\documentclass[aps,prl,reprint,showpacs,superscriptaddress]{revtex4-1}
\usepackage[plainpages=false,pdfpagelabels,colorlinks=true,linkcolor=red,urlcolor=blue,citecolor=blue,pdftitle={Title},pdfauthor={},pdfdisplaydoctitle=true,pdfduplex=DuplexFlipLongEdge]{hyperref}
\usepackage{verbatim}
\usepackage[english]{babel}
\usepackage{bm}
\usepackage{amsmath}
\usepackage{epsfig}
\usepackage{array}
\usepackage{color, soul}
\usepackage{braket}

\begin{document}

\title{Realizing a quantum generative adversarial network \\ using a programmable superconducting processor}

\author{Kaixuan~Huang}
	\thanks{These authors contributed equally to this work.}
	\affiliation{The Key Laboratory of Weak Light Nonlinear Photonics, Ministry of Education, Teda Applied Physics Institute and School of Physics, Nankai University, Tianjin 300457, China}

\author{Zheng-An~Wang}
	\thanks{These authors contributed equally to this work.}
	\affiliation{Institute of Physics, Chinese Academy of Sciences, Beijing 100190, China}

\author{Chao~Song}
	\thanks{These authors contributed equally to this work.}
	\affiliation{Interdisciplinary Center for Quantum Information, State Key Laboratory  of Modern Optical Instrumentation,
		and Zhejiang Province Key Laboratory of Quantum Technology and Device,\\
		Department of Physics, Zhejiang University, Hangzhou 310027, China}

\author{Kai~Xu}
	\affiliation{Institute of Physics, Chinese Academy of Sciences, Beijing 100190, China}

\author{Hekang~Li}
	\affiliation{Interdisciplinary Center for Quantum Information, State Key Laboratory  of Modern Optical Instrumentation,
		and Zhejiang Province Key Laboratory of Quantum Technology and Device,\\
		Department of Physics, Zhejiang University, Hangzhou 310027, China}

\author{Zhen~Wang}
\affiliation{Interdisciplinary Center for Quantum Information, State Key Laboratory  of Modern Optical Instrumentation,
	and Zhejiang Province Key Laboratory of Quantum Technology and Device,\\
	Department of Physics, Zhejiang University, Hangzhou 310027, China}

\author{Qiujiang~Guo}
	\affiliation{Interdisciplinary Center for Quantum Information, State Key Laboratory  of Modern Optical Instrumentation,
		and Zhejiang Province Key Laboratory of Quantum Technology and Device,\\
		Department of Physics, Zhejiang University, Hangzhou 310027, China}
\author{Zixuan~Song}
\affiliation{Interdisciplinary Center for Quantum Information, State Key Laboratory  of Modern Optical Instrumentation,
	and Zhejiang Province Key Laboratory of Quantum Technology and Device,\\
	Department of Physics, Zhejiang University, Hangzhou 310027, China}

\author{Zhi-Bo~Liu}
	\email{liuzb@nankai.edu.cn}
	\affiliation{The Key Laboratory of Weak Light Nonlinear Photonics, Ministry of Education, Teda Applied Physics Institute and School of Physics, Nankai University, Tianjin 300457, China}

\author{Dongning~Zheng}
	\affiliation{Institute of Physics, Chinese Academy of Sciences, Beijing 100190, China}
	\affiliation{CAS Center for Excellence in Topological Quantum Computation, University of Chinese Academy of Sciences, Beijing 100190, China}

\author{Dong-Ling~Deng}
\email{dldeng@tshinghua.edu.cn}
	\affiliation{Center for Quantum Information, Institute for Interdisciplinary Information Sciences, Tsinghua University, Beijing 100084, China}

\author{H.~Wang}
	\affiliation{Interdisciplinary Center for Quantum Information, State Key Laboratory  of Modern Optical Instrumentation,
		and Zhejiang Province Key Laboratory of Quantum Technology and Device,\\
		Department of Physics, Zhejiang University, Hangzhou 310027, China}
\author{Jian-Guo~Tian}
\affiliation{The Key Laboratory of Weak Light Nonlinear Photonics, Ministry of Education, Teda Applied Physics Institute and School of Physics, Nankai University, Tianjin 300457, China} 

\author{Heng~Fan}
    \email{hfan@iphy.ac.cn}
    \affiliation{Institute of Physics, Chinese Academy of Sciences, Beijing 100190, China}
    \affiliation{School of Physical Sciences, University of Chinese Academy of Sciences, Beijing 100190, China}
		
\begin{abstract}
{\bf  Generative adversarial networks are an emerging technique with wide applications in machine learning ~\cite{Goodfellow2014Generative}, which have achieved dramatic success in a number of challenging tasks including image and video generation ~\cite{Creswell2018Generative}. 
When equipped with quantum processors, their quantum counterparts---called quantum generative adversarial networks (QGANs)---may even exhibit exponential advantages in certain machine learning applications \cite{Lloyd2018Quantum, Demers2018Quantum,Zeng2019Learning,Zoufal2019Quantum,hu2019quantum,regitti2019,regitti2020}. Here, we report an experimental implementation of a QGAN using a programmable superconducting processor, in which both the generator and the discriminator are parameterized via layers of single- and multi-qubit quantum gates. The programmed QGAN runs automatically several rounds of adversarial learning with quantum gradients to achieve a Nash equilibrium point, where the generator 
can replicate data samples that mimic the ones from the training set. Our implementation is promising to scale up to noisy intermediate-scale quantum devices, thus paving the way for experimental explorations of quantum advantages in practical applications with near-term quantum technologies.
}
\end{abstract}

\maketitle


The interplay between quantum physics and machine learning gives rise to an emergent research frontier of quantum machine learning that has attracted tremendous attention recently \cite{Carleo2019Machine,Sarma2019Machine,Biamonte2017Quantum,Dunjko2018Machine}.
In particular, certain carefully-designed quantum algorithms for machine learning, or more broadly artificial intelligence, may exhibit exponential advantages compared to their best possible classical counterparts \cite{Sarma2019Machine,Biamonte2017Quantum,Dunjko2018Machine,gao2018quantum,harrow2019low,Lloyd2018Quantum,Havlicek2019}.
An intriguing example concerns quantum generative adversarial networks (QGANs) \cite{Lloyd2018Quantum},
where near-term quantum devices have the potential to showcase quantum supremacy \cite{arute2019quantum} with real-life practical applications.

The general framework of QGANs consists of a generator learning to generate statistics for data mimicking those of a true data set, and a discriminator trying to discriminate generated data from true data \cite{Lloyd2018Quantum}. The generator and  discriminator follow an adversarial learning procedure to optimize their strategies alternatively and arrive at a Nash equilibrium point, where the generator learns the underlying statistics of the true data and the discriminator can no longer distinguish the difference between the true and generated data.
Different versions of QGANs have been proposed~\cite{Demers2018Quantum,Lloyd2018Quantum,Zeng2019Learning,Zoufal2019Quantum,hu2019quantum,regitti2019,regitti2020}. 
Recently, a proof-of-principle experimental QGAN demonstration has been reported \cite{hu2019quantum}, showing that the generator can indeed be trained via the adversarial learning process to replicate the statistics of the single-qubit quantum data output from a quantum channel simulator. Yet, in this experiment the gradient, which is crucial for training the QGAN, was estimated numerically by the finite-difference method with a classical computer. This finite-difference approach induces an inaccuracy to the gradient and therefore may significantly retard the convergence to the equilibrium point \cite{harrow2019low}. In addition, the quantum states involved in this experiment were all single-qubit states, and entanglement, a characterizing feature of quantumness and a vital resource for quantum advantages, was absent during the learning process \cite{Biamonte2017Quantum}. 

In this paper, we add these two crucial yet missing blocks by reporting an experiment realization of a QGAN based on a programmable superconducting processor with multiple qubits. 
Superconducting qubits are a promising platform for realizing QGANs, owing to their flexible design, excellent scalability and remarkable controllability. 
In our implementation, both the generator and discriminator are composed of multiqubit parameterized quantum circuits, also referred to as quantum neural networks in some contexts~\cite{Jarrod2018QNN, Cerezo2020Plateaus, Andrea2020QNN, Patrick2020}.
Here, we benchmark the functionality of the quantum gradient method by learning an arbitrary mixed state, where the state is replicated with a fidelity up to 0.999. We further utilize our QGAN to learn an classical XOR gate, and the generator is successfully trained to exhibit a truth table close to that of the XOR gate.
\begin{figure*}
	\includegraphics[width=1.0\textwidth]{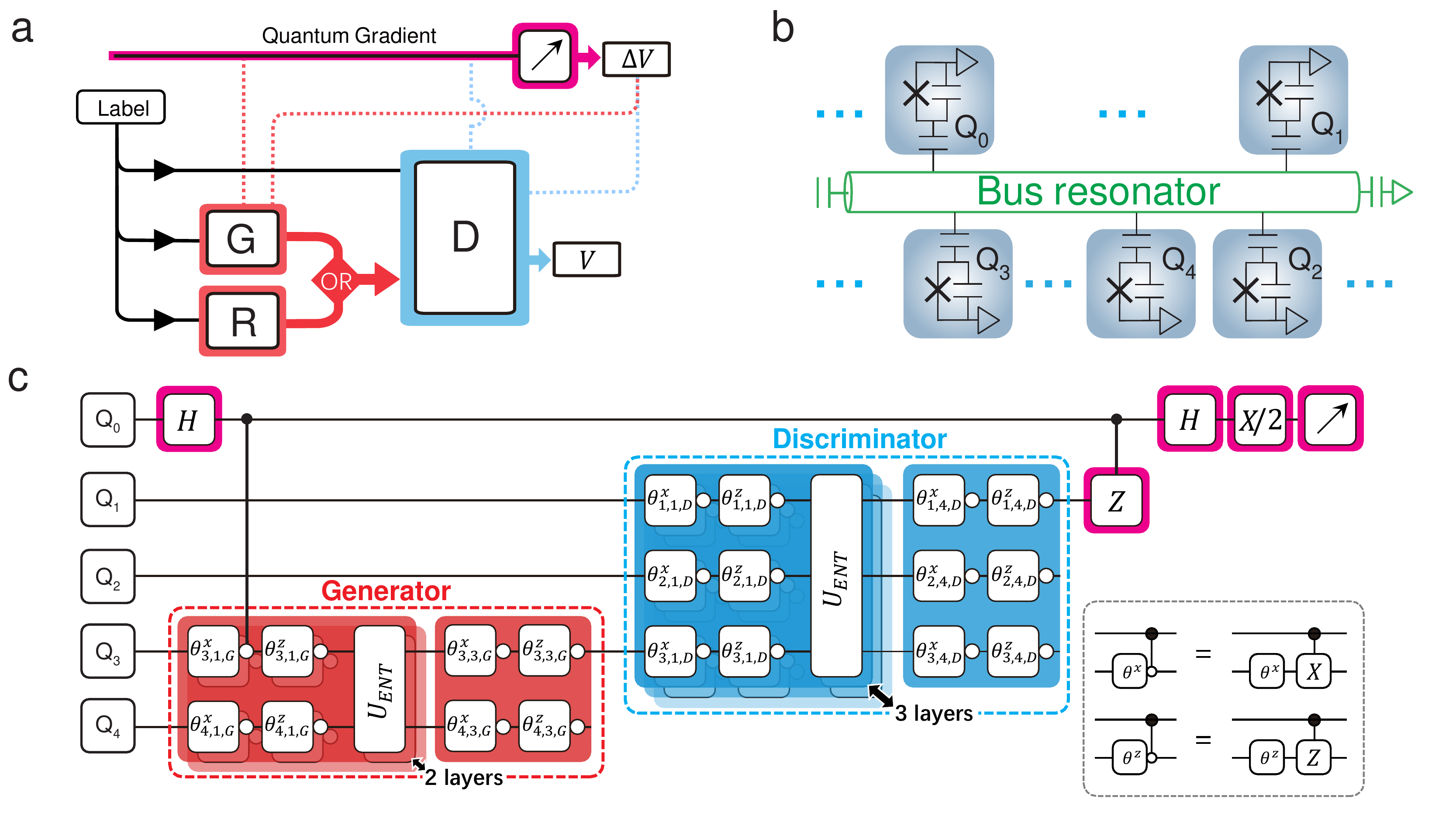}
	\caption{\textbf{QGAN algorithm and its implementation.} \textbf{a,} Overview of the QGAN logic. 
		\textbf{b,} Sketch of the superconducting processor used to implement the QGAN algorithm, where the five qubits, Q$_0$ to Q$_4$, are interconnected by the central bus resonator. 
		\textbf{c,}
		An instance of the experimental sequences for fulfilling the QGAN algorithm to learn the classical XOR gate. Both \textbf{G} and \textbf{D} are parameterized quantum circuits consisting of layers of the multiqubit entangling gate $U_{\textrm{ENT}}$ and the single-qubit rotations $\theta_{j,l,m}^{x/z}$, where $l$ is the layer index, $m \in \{\textbf{G},\textbf{D}\}$, and the superscript (x or z) refers to the axis in the Bloch sphere around which the state of Q$_j$ is rotated by the angle $\theta$.
		Shown is the training sequence on \textbf{G} to optimize its parameter $\theta_{3,1,G}^x$ based on the quantum gradient subroutine, which includes two Hadamard gates ($H$), two controlled rotation gates, 
		and a $\pi/2$ rotation around x-axis ($X/2$).
		In this instance, both $Q_1$-$Q_2$ and $Q_3$-$Q_4$ store the input label, 
		and \textbf{D}'s output score is encoded in Q$_1$ by $S^{{\textrm{D}},\textrm{R/G}}_n = \langle\sigma^z_{1}\rangle/2+1/2$, which can be obtained by directly measuring Q$_1$.
		$Q_0$'s $\langle\sigma_0^z\rangle$ gives the score derivative with respect to the rotational angle parameter right before the first controlled rotation gate, i.e., $\partial\langle\sigma_1^z\rangle / \partial \theta_{3,1,G}^x = -\langle\sigma_0^z\rangle$ for the sequence displayed here. The first controlled rotation gate can be either a controlled X (CNOT) or controlled Z (CZ) gate depending on the single-qubit rotation axis it follows as shown in the lower right box.
	}
	\label{fig:Fig1}
\end{figure*}

We first introduce a general recipe for our QGAN and then apply it to two typical scenarios: learning quantum states and the classical XOR gate. 
The overall structure of the QGAN is outlined in Fig.~\ref{fig:Fig1} (a), which includes 
an input label, a real source (\textbf{R}), a generator (\textbf{G}), a discriminator (\textbf{D}), and a quantum gradient subroutine. 
The input label sorts the training data stored in \textbf{R}, and also instructs \textbf{G} to generate data samples mimicking \textbf{R}. \textbf{D} receives the label and corresponding data samples from either \textbf{G} or \textbf{R}, and then evaluates with appropriate scores, 
based on which a loss function $V$ is constructed to differentiate between \textbf{R} and \textbf{G}. 
The adversarial training procedure is repeated in conjugation with the quantum gradient subroutine, which yields the partial derivatives of $V$ with respect to the parameters constructing \textbf{D} or \textbf{G}, so as to maximize $V$ for the optimal configuration of \textbf{D} and to minimize $V$ for the optimal \textbf{G} in terms of the chosen values of the constructing parameters.

Specifically, denoting the sets of parameters constructing \textbf{G} and \textbf{D} as $\vec{\theta}_\textrm{G}$ and $\vec{\theta}_\textrm{D}$, respectively,
the loss function $V$ is written as
\begin{equation}
V=\frac{1}{N} \sum_{n=1}^{N} \left[S_{n}^{\textrm{D}, \textrm{R}}(\vec{\theta}_\textrm{D})-S_{n}^{\textrm{D},\textrm{G}}(\vec{\theta}_\textrm{D},\vec{\theta}_\textrm{G})\right], \nonumber
\label{loss}
\end{equation}
where $N$ is the total number of data samples selected for training and
$S_{n}^{\textrm{D}, \textrm{R}}$ ($S_{n}^{\textrm{D}, \textrm{G}}$) represents the score of the $n$th data sample from \textbf{R} (\textbf{G}) evaluated by \textbf{D}. 
\textbf{G} and \textbf{D} are trained alternately with \textbf{D} being trained first.
In \textbf{D}'s turn, we maximize the loss function by iteratively optimizing $\vec{\theta}_\textrm{D}$ according to
$\vec{\theta}^{i+1}_\textrm{D}=\vec{\theta}^i_\textrm{D}+ \alpha_\textrm{D} \nabla_{\vec{\theta}_\textrm{D}}V$,
where $i$ is the iteration step index and $\nabla_{\vec{\theta}_\textrm{D}}V$ denotes the gradient vector of the loss function 
at the $i$th step;
we train \textbf{G} by minimizing the square of the loss function with an iteration relation 
$\vec{\theta}^{i+1}_\textrm{G}=\vec{\theta}^i_\textrm{G}-\alpha_\textrm{G} \nabla_{\vec{\theta}_\textrm{G}}V^2$. 
The learning rates can be adjusted by tuning $\alpha_\textrm{D}$ and $\alpha_\textrm{G}$ which are typically on the order of unity.

An important quantity that plays a vital role in training our QGAN is the gradient of the loss function with respect to a given parameter. Interestingly, owing to the special structures of our quantum circuits for the generator and the discriminator, a common approach to obtain such a gradient is to shift the corresponding parameter by $\pm \pi/2$ and measure the loss function at the shifted values \cite{Schuld2019}. In our experiment,
we employ a even simpler method---the Hadamard test quantum algorithm \cite{Mitarai2019}---to obtain the gradient.
This algorithm can reduce half of the time in comparison with the common approach to finish the training,
at the expense of an additional auxiliary qubit and two controlled gates (see next) \cite{SM}. 
At each round of the training process, the parameters of the discriminator (or generator) are updated simultaneously after the gradients for all parameters are obtained, and a quantum process tomography (QPT) is performed to characterize the overlap fidelity between the generated data based on the updated parameters and the true data.

\begin{figure}[htb]
	\includegraphics[width=0.49\textwidth]{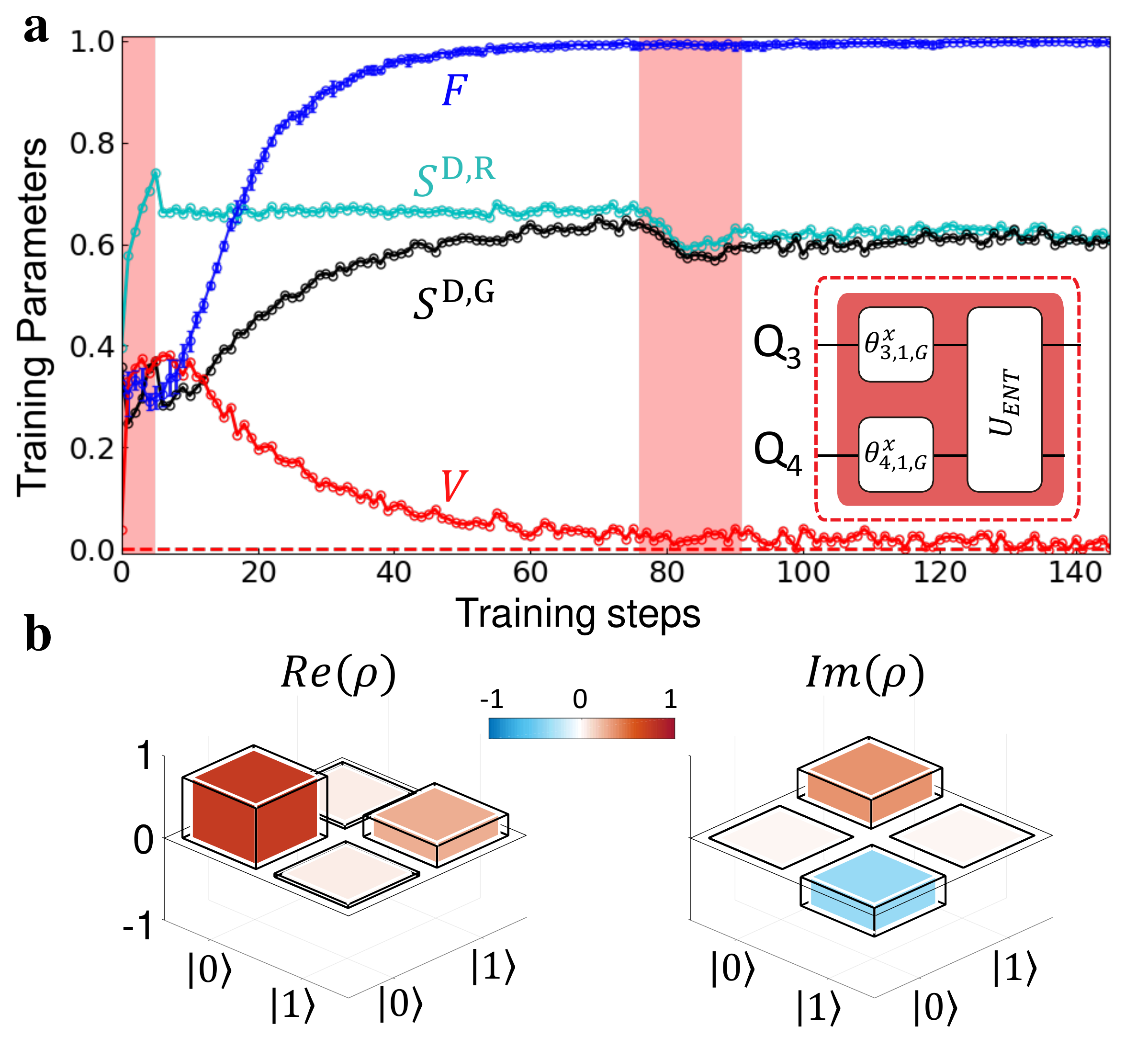}
	\caption{{\bf QGAN performance in learning a mixed state.} \textbf{a,} Tracking of  the loss function $V$, the output scores $S^{\textrm{D}, \textrm{R/G}}$, and the state fidelity between \textbf{R}/\textbf{G}'s output states $F$ during the adversarial training procedure with learning rates $\alpha_\textrm{D} = 0.8$ and $\alpha_\textrm{G} = 0.6$. The alternate training stages of \textbf{D} and \textbf{G} are marked by red and white regions, respectively. In each stage, the maximum step number is limited to 50 for \textbf{D} and 100 for \textbf{G}. Inset: the quantum circuit for generating $\rho_\textrm{R}$ for Q$_3$. The same circuit is also used for \textbf{G}, with random initial guesses for the two rotational angles which are then optimized.
	\textbf{b,} The real and imaginary parts of the output density matirces of \textbf{R} (black frames) and the final \textbf{G} (solid bars).
	}
	\label{fig:Fig2}
\end{figure}

The above mentioned QGAN is experimentally realized on a superconducting quantum processor using 5 frequency-tunable transmon qubits labeled as Q$_j$ for $j=0$ to 4, where
all qubits are interconnected by a central bus resonator as illustrated in Fig.~\ref{fig:Fig1} (b).
The role arrangement of Q$_0$ to Q$_4$ can be visualized by the exemplary experimental sequence shown in Fig~\ref{fig:Fig1} (c). Q$_0$ is to assist the quantum gradient subroutine. Q$_1$-Q$_2$ stores the label which is passed to \textbf{D}, with the output score encoded in Q$_1$ by $S^{{\textrm{D}},\textrm{R/G}}_n = \langle\sigma^z_{1}\rangle/2+1/2$. For the experimental instances with \textbf{G} inserted, Q$_3$-Q$_4$ also stores the label as the input to \textbf{G} and the output data sample from \textbf{G} is passed to \textbf{D} via Q$_3$; for the experimental instances with \textbf{R} in replacement of \textbf{G}, only Q$_3$ stores the data sample from \textbf{R} as designated by the label. Details of the device parameters can be found in supplementary materials (SM)~\cite{SM} and Ref.~\cite{Song2019Generation}.


By tuning the qubits on resonance but detuned from the bus resonator in frequency, these qubits can be all effectively  connected, which enables the flexible realizations of the multiqubit entangling gates among arbitrarily selected qubits.
The all-to-all interactions are described in the dispersive regime by the effective Hamiltonian $H_{\textrm{I}}=\sum \lambda_{jk}(\sigma_j^+\sigma_k^-+\sigma_j^-\sigma_k^+)$~\cite{Song2019Generation}, where $\sigma_j^+$ ($\sigma_j^-$) denotes the raising (lowering) operator of Q$_j$, and $\lambda_{jk}$ is the effective coupling strength between Q$_j$ and Q$_k$ mediated by the bus resonator.
Evolution under this Hamiltonian for an interaction time $\tau$ leads to the entangling operator with the form of $U_{\textrm{ENT}}=e^{-iH_{\textrm{I}}\tau}$, which can steer the interacting qubits into highly entangled state.
In our QGAN, the parameterized quantum circuits that comprise \textbf{G} and \textbf{D} leverage the naturally available multiqubit $U_{\textrm{ENT}}$s, with
the interaction time of the two-qubit $U_{\textrm{ENT}}$ fixed at around 50~ns for \textbf{G} and the three-qubit one fixed at around 55~ns for \textbf{D}~\cite{SM}.
We note that, in this hardware-efficient realization of the QGAN, the entangling operators can be any device-tailored operations that generate sufficient entanglement.

As laid out in Fig.~\ref{fig:Fig1} (c), the entangling operators $U_{\textrm{ENT}}$ are interleaved with the single-qubit $X$ and $Z$ rotations, 
which successively rotate Q$_j$ around x- and z-axis in the Bloch sphere by angles of $\theta_{j,l,m}^x$ and $\theta_{j,l,m}^z$, where 
$l$ is the layer index and $m \in \{\textbf{G}, \textbf{D}\}$. 
The lengths of the $X$ and $Z$ rotations are fixed at 30 and 20 ns, respectively.
Taking into account the experimental imperfections, we perform numerical simulations to decide the depths of the interleaved layers consisting of $U_{\textrm{ENT}}$ and the single-qubit rotations,
for a balance between the learning fidelity and efficiency.
For example, to learn the XOR gate, the circuit layer depths are set to be 2 and 3 for \textbf{G} and \textbf{D} respectively, as shown in Fig.\ref{fig:Fig1} (c).

The QGAN learning process is guided by the gradient of the loss function with respect to $\vec{\theta}_\textrm{G}$ and $\vec{\theta}_\textrm{D}$.
To obtain these gradients, we adopt a quantum method called Hadamard test~\cite{Demers2018Quantum, Mitarai2019}, 
which is illustrated in the sequence instance in Fig.~\ref{fig:Fig1} (c).
For the partial derivative with respect to the parameter $\theta_{j,l,m}^{x}$ ($\theta_{j,l,m}^{z}$),
we insert the first controlled-X (Z) gate right after the single-qubit X (Z) rotation containing this parameter, with Q$_j$ as the target.
The second controlled-Z gate is applied at the end of the training sequence with Q$_1$ as the target.
The partial derivative of Q$_1$'s $\langle\sigma^z_{1}\rangle$, which relates to \textbf{D}'s output score, 
is given by $\partial\langle\sigma^z_{1}\rangle/\partial\theta_{j,l,m}^{x (z)}=-\langle\sigma^z_{0}\rangle$, which can be directly obtained by measuring Q$_0$ and used in computing the gradient of the loss function.
See SM~\cite{SM} for more
details about the experimental realizations of the controlled-X (Z) gates,
as well as the theoretical and experimental verifications of the quantum gradient method.

To benchmark the functionality of the quantum gradient method and the learning efficiency of our QGAN circuit, 
we first train an arbitrary mixed state as data which is a simulation of quantum channel. 
As shown in the inset of Fig.\ref{fig:Fig2} (a), the mixed state for Q$_3$ reads
$\rho_\textrm{R} = \left(\begin{smallmatrix} 0.7396 &0.0431 + 0.3501i\\0.0431 - 0.3501i&0.2604\protect \end{smallmatrix}\right)$,
which is generated by applying two single-qubit X rotations (with the rotation angles of 1.35 on Q$_3$ and 0.68 on Q$_4$) followed by the 
$U_{\textrm{ENT}}$ gate on Q$_3$ and Q$_4$.
Correspondingly, \textbf{G} is set up with a single layer and 2 parameters describing the single-qubit X rotation angles during the training, while \textbf{D} remains the one shown in Fig.\ref{fig:Fig1} (c) with 3 layers and all 18 parameters being trained. 
The trajectories of the loss function and scores of data from \textbf{R}/\textbf{G} during the training process are recorded and plotted in Fig.~\ref{fig:Fig2} (a). 
We optimize \textbf{D} at the beginning of the training to enlarge the distance between $S^{\textrm{D}, \textrm{R}}$ and $S^{\textrm{D}, \textrm{G}}$. At the end of this turn, \textbf{D} can discriminate datasets from \textbf{R} and \textbf{G} with the maximum probability.
In \textbf{G}'s turn, 
$S^{\textrm{D}, \textrm{G}}$ moves towards $S^{\textrm{D}, \textrm{R}}$ which means that \textbf{G} is learning the behavior of \textbf{R}. Each turn ends when the optimal point of the loss function is reached, or the iteration number 
goes over a preset limit.
As the adversarial learning process goes on, the value of the loss function oscillates from turn to turn 
and eventually converges to 0 
indicating that the learning 
arrives at a Nash equilibrium point, where \textbf{G} is able to produce a mixed state $\rho_\textrm{G}$ which resembles $\rho_\textrm{R}$
and \textbf{D} can no longer distinguish between them~\cite{Lloyd2018Quantum,Demers2018Quantum}.
The training process is characterized by the similarity between datasets generated by \textbf{G} and \textbf{R}, which is quantified by the state fidelity $F(\rho_{\textrm{R}}, \rho_{\textrm{G}}) = \textrm{Tr}(\sqrt{\rho_{\textrm{R}}}\rho_{\textrm{G}}\sqrt{\rho_{\textrm{R}}})$, 
As shown in Fig~\ref{fig:Fig2} (a), $F$ increases rapidly with the iteration steps, indicating the effectiveness of the adversarial learning.
The density matrices $\rho_{\textrm{R}}$ and the final $\rho_{\textrm{G}}$ are plotted in Fig~\ref{fig:Fig2} (b), which yields a state fidelity of around 0.999. 

\begin{figure}
	\includegraphics[width=0.49\textwidth]{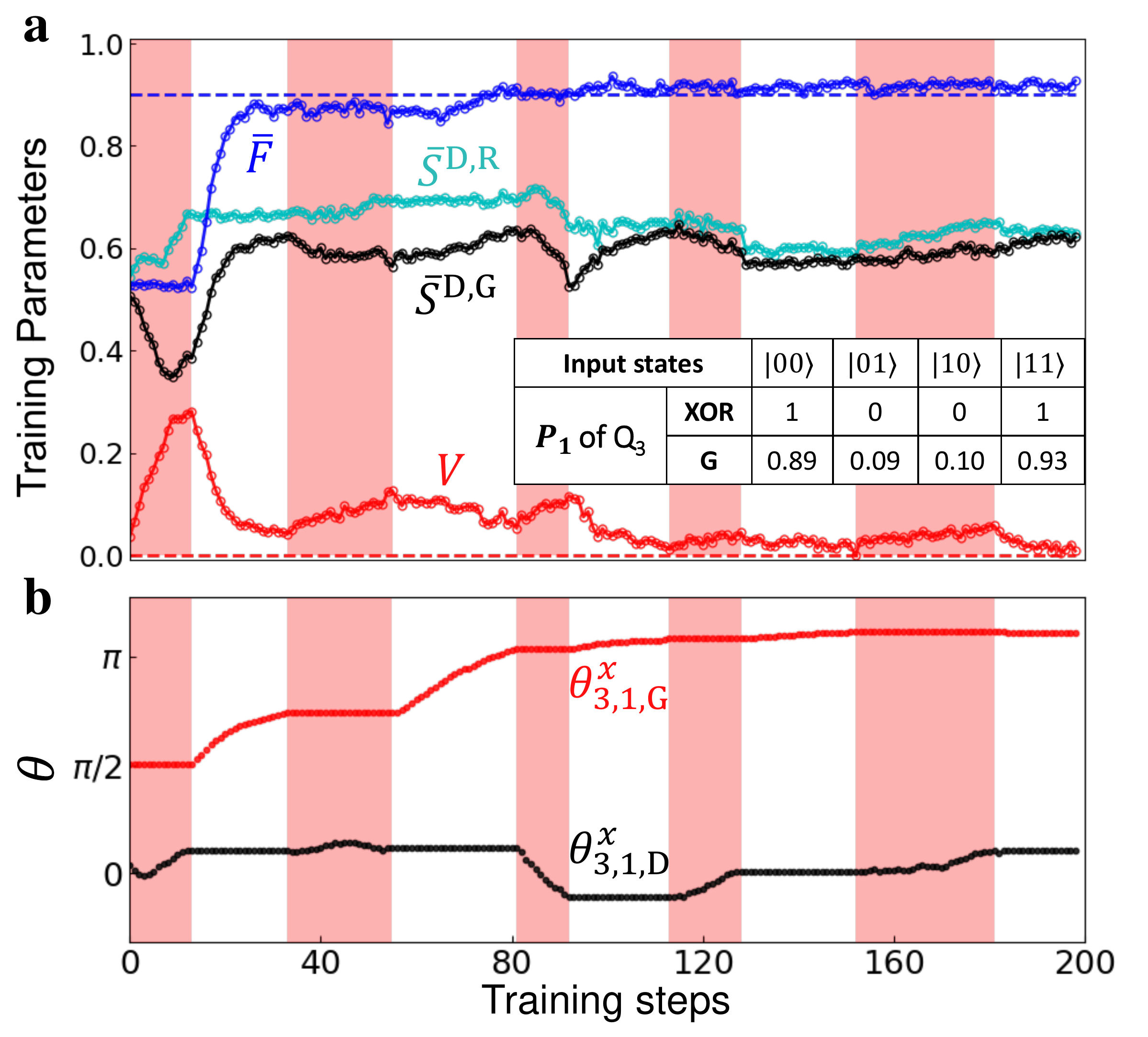}
	\caption{{\bf QGAN performance in learning the XOR gate}. \textbf{a,} Tracking of  of the loss function $V$, the averaged output scores $\bar{S}^{\textrm{D}, \textrm{R/G}}$, and the averaged state fidelity between \textbf{R}/\textbf{G}'s output states $\bar{F}$ during the adversarial training process with learning rates $\alpha_\textrm{D} = 1.0$ and $\alpha_\textrm{G} = 1.5$. The maximum step number is limited to 50 for both $\textbf{G}$ and $\textbf{D}$. $\bar{S}^{\textrm{D}, \textrm{R/G}}$ and $\bar{F}$ are averaged over four possible inputs. 
	Inset: The truth table of \textbf{G} after training in comparison with that of the XOR gate.
		\textbf{b,} Trajectories of two representative parameters constructing \textbf{D} and \textbf{G}, $\theta_{3,1,D}^x$ and $\theta_{3,1,G}^x$, during the QGAN training. 
		All parameters in \textbf{G} and \textbf{D} are initialized randomly between 0 and $\pi$ and optimized alternately during the training.
	}
	\label{fig:Fig3}
\end{figure}

Now we apply the recipe to train the QGAN to replicate the statistics of an XOR gate, which is a classical gate with input-output rules $00 \rightarrow 0$, $01 \rightarrow 1$, $10 \rightarrow 1$, and $11 \rightarrow 0$. We use the computational basis state $|0\rangle$ and $|1\rangle$ to encode the classical data $0$ and $1$, respectively. 
We randomly initialize the parameters of both \textbf{D} and \textbf{G}, and then update them alternately following the rules outlined  above.
The trajectories of the key parameters benchmarking the QGAN performance
during the training process are 
plotted in Fig.~\ref{fig:Fig3} (a), with the evolutions of two representative parameters construcing \textbf{D} and \textbf{G} shown in Fig.~\ref{fig:Fig3} (b).
Again, the loss function exhibits a typical oscillation during the adversarial process, and the training reaches its equilibrium after about 190 steps with an average state fidelity of 0.927. In addition, after training \textbf{G} successfully exhibits a truth table close to that of an XOR gate, as shown in the inset of Fig.~\ref{fig:Fig3} (a). 

In conclusion, we have experimentally implemented a multi-qubit QGAN equipped with a quantum gradient algorithm on a programmable superconducting processor.
The results clearly show
the feasibility of QGAN for learning data with both classical and quantum statistics.
The parameterized quantum circuits for constructing quantum generators and discriminators do not require accurate implementations of specific quantum logics and can be achieved on the near-term quantum devices across different physical platforms.
Our implementation paves the way to the much-anticipated computing paradigm with combined quantum-classical processors, and holds the intriguing potential to realize practical quantum supremacy \cite{arute2019quantum} with noisy intermediate-scale quantum devices
\cite{Preskill2018quantumcomputingin}.


\section*{Acknowledgments}
\noindent Devices were made at the Nanofabrication Facilities at
Institute of Physics in Beijing and National Center for Nanoscience
and Technology in Beijing. The experiment was performed on the quantum computing platform at Zhejiang University.
{\bf Funding:} Supported by 
National Basic Research Program of China (Grants No. 2017YFA0304300, No. 2016YFA0302104 and No.
2016YFA0300600),
National Natural Science Foundation of China (Grants No. 11934018 and No. 11725419), 
the Zhejiang Province Key Research and Development Program (Grant No. 2020C01019),
the start-up fund from Tsinghua University (Grant No. 53330300320), and 
Strategic Priority Research Program of Chinese Academy of Sciences 	
(Grant No. XDB28000000).
{\bf Author contributions:} Z.A.W, D.L.D. and H.F. proposed the idea; supervised by Z.B.L, J.G.T and H.W., 
K.H., C.S., K.X. and Q.G. conducted the experiment;
supervised by H.F., K.H. and Z.A.W. performed the numerical simulation; 
H.L. and D.Z. fabricated the device;
K.H., Z.A.W., C.S., Z.W., D.L.D., H.W. and H.F. cowrote the manuscript; and all authors contributed to the experimental setup, discussions of the results, and development of the manuscript.
{\bf Competing interests:} Authors declare no competing interests.
{\bf Data and materials availability:} All data needed to evaluate the
 conclusions in the paper are present in the paper or the
 supplementary materials.


\begin{thebibliography}{}

\bibitem{Goodfellow2014Generative} I. Goodfellow, J. Pouget-Abadie, M. Mirza, B. Xu, D. Warde-Farley, S. Ozair, A. Courville, and Y. Bengio, 
in \textit{Advances in neural information processing systems} (2014) pp. 2672–2680.

\bibitem{Creswell2018Generative} A. Creswell, T. White, V. Dumoulin, K. Arulkumaran, B. Sengupta, and A. A. Bharath, 
\textit{IEEE Signal Processing Magazine} {\bf 35}, 53 (2018).

\bibitem{Lloyd2018Quantum} S. Lloyd and C. Weedbrook, 
\textit{Phys. Rev. Lett.} {\bf 121}, 040502(2018).

\bibitem{Demers2018Quantum} P.-L. Dallaire-Demers and N. Killoran, 
\textit{Phys. Rev. A} {\bf 98}, 012324 (2018).

\bibitem{Zeng2019Learning} J. Zeng, Y. Wu, J.-G. Liu, L. Wang, and J. Hu, 
\textit{Phys. Rev. A} {\bf 99}, 052306 (2019).

\bibitem{Zoufal2019Quantum} C. Zoufal, A. Lucchi, and S. Woerner, 
\textit{npj Quantum Information} {\bf 5}, 103 (2019).

\bibitem{hu2019quantum} L. Hu, S.-H. Wu, W. Cai, Y. Ma, X. Mu, Y. Xu, H. Wang, Y. Song, D.-L. Deng, C.-L. Zou, and L. Sun, 
\textit{Science advances} {\bf 5}, eaav2761 (2019).

\bibitem{regitti2019} J. Romero and A. Aspuru-Guzik, 
\textit{arXiv:1901.00848} (2019).

\bibitem{regitti2020} A. Anand, J. Romero, M. Degroote, and A. Aspuru-Guzik, 
\textit{arXiv:2006.01976} (2020).

\bibitem{Carleo2019Machine} G. Carleo, I. Cirac, K. Cranmer, L. Daudet, M. Schuld, N. Tishby, L. Vogt-Maranto, and L. Zdeborová, 
\textit{Rev. Mod. Phys.} {\bf 91}, 045002 (2019).

\bibitem{Sarma2019Machine} S. D. Sarma, D.-L. Deng, and L.-M. Duan, 
\textit{Physics Today} {\bf 72}, 48 (2019).

\bibitem{Biamonte2017Quantum} J. Biamonte, P. Wittek, N. Pancotti, P. Rebentrost, N. Wiebe, and S. Lloyd, 
\textit{Nature} {\bf 549}, 195 (2017).

\bibitem{Dunjko2018Machine} V. Dunjko and H. J. Briegel, 
\textit{Rep. Prog. Phys.} {\bf 81}, 074001 (2018).

\bibitem{gao2018quantum} X. Gao, Z.-Y. Zhang, and L.-M. Duan, 
\textit{Science advances} {\bf 4}, eaat9004 (2018).

\bibitem{harrow2019low} A. Harrow and J. Napp,
 \textit{arXiv:1901.05374} (2019). 

\bibitem{Havlicek2019} V. Havlicek, A. D. Corcoles, K. Temme, A. W. Harrow,  A. Kandala, J. M. Chow, and J. M. Gambetta, 
\textit{Nature} {\bf 567}, 209 (2019).

\bibitem{arute2019quantum} F. Arute, K. Arya, R. Babbush, D. Bacon, J. C. 
Bardin, R. Barends, R. Biswas, S. Boixo, F. G. S. L.
Brandao, D. A. Buell, B. Burkett, Y. Chen, Z. Chen,
B. Chiaro, R. Collins, W. Courtney, A. Dunsworth,
E. Farhi, B. Foxen, A. Fowler, C. Gidney, M. Giustina,
R. Graff, K. Guerin, S. Habegger, M. P. Harrigan,
M. J. Hartmann, A. Ho, M. Hoffmann, T. Huang,
T. S. Humble, S. V. Isakov, E. Jeffrey, Z. Jiang,
D. Kafri, K. Kechedzhi, J. Kelly, P. V. Klimov, S. Knysh,
A. Korotkov, F. Kostritsa, D. Landhuis, M. Lindmark,
E. Lucero, D. Lyakh, S. Mandrà, J. R. Mc-
Clean, M. McEwen, A. Megrant, X. Mi, K. Michielsen,
M. Mohseni, J. Mutus, O. Naaman, M. Neeley, C. Neill,
M. Y. Niu, E. Ostby, A. Petukhov, J. C. Platt, C. Quintana,
E. G. Rieffel, P. Roushan, N. C. Rubin, D. Sank,
K. J. Satzinger, V. Smelyanskiy, K. J. Sung, M. D. Trevithick,
A. Vainsencher, B. Villalonga, T. White, Z. J.
Yao, P. Yeh, A. Zalcman, H. Neven, and J. M. Martinis, 
\textit{Nature} {\bf 574}, 505 (2019).

\bibitem{Jarrod2018QNN} M. Jarrod, R., B. Sergio, S. Vadim, N., B. Ryan, and N. Hartmut, 
 \textit{Nat. Commun.} {\bf 9} (2018), 10.1038/s41467-018-07090-4.

\bibitem{Cerezo2020Plateaus} M. Cerezo, A. Sone, T. Volkoff, L. Cincio, and P. J. Coles, 
 \textit{arXiv:2001.00550} (2020).

\bibitem{Andrea2020QNN} A. Skolik, J. R. McClean, M. Mohseni, P. Smagt, and M. Leib, 
 \textit{arXiv:2006.14904} (2020). 

\bibitem{Patrick2020} P. Huembeli and A. Dauphin, 
 \textit{arXiv:2008.02785} (2020).

\bibitem{Schuld2019} M. Schuld, V. Bergholm, C. Gogolin, J. Izaac, and N. Killoran, 
\textit{Phys. Rev. A} {\bf 99}, 032331 (2019).

\bibitem{Mitarai2019} K. Mitarai and K. Fujii, 
\textit{Physical Review Research} {\bf 1}, 013006 (2019).
 
 
\bibitem{SM} Supplementary-Materials.

\bibitem{Song2019Generation} C. Song, K. Xu, H. Li, Y.-R. Zhang, X. Zhang, W. Liu,
Q. Guo, Z. Wang, W. Ren, J. Hao, H. Feng, H. Fan, D. Zheng, D.-W. Wang, H. Wang, and S.-Y. Zhu, 
\textit{Science} {\bf 365}, 574 (2019).

\bibitem{Preskill2018quantumcomputingin} J. Preskill, 
 \textit{Quantum} {\bf 2}, 79 (2018).
\end{thebibliography}

\begin{thebibliography}{}

\bibitem{song2019} C. Song, K. Xu, H. Li, Y.-R. Zhang, X. Zhang, W. Liu, Q. Guo, Z. Wang, W. Ren, J. Hao, H. Feng, H. Fan, D. Zheng, D.-W. Wang, H. Wang, and S.-Y. Zhu
{\textit{Science} \textbf{365}, 574-577 (2019).}

	
\bibitem{song2017}  C. Song, K. Xu, W. Liu, C.-P Yang, S.-B. Zheng, H. Deng, Q. Xie, K. Huang,  Guo, L. Zhang, P. Zhang, D. Xu, D. Zheng, X. Zhu, H. Wang, Y.-A. Chen, C.-Y. Lu, S. Han, and J-W. Pan,  
 {\textit{Phys. Rev. Lett.} \textbf{119}, 180511 (2017).}

\bibitem{Motzoi2009}F. Motzoi, J. M. Gambetta, P. Rebentrost, and F. K. Wilhelm.
{\textit{Phys. Rev. Lett.} \textbf{103}, 110501 (2009).}

\bibitem{qguo2018} Q. Guo, S.-B. Zheng, J. Wang, C. Song, P. Zhang, K. Li, W. Liu, H. Deng, K. Huang, D. Zheng, X. Zhu, H. Wang, C.-Y. Lu, and J.-W. Pan, 
 {\textit{Phys. Rev. Lett.} \textbf{121}, 130501 (2018).}

\bibitem{NKpra2018}P.-L. Dallaire-Demers, N. Killoran. 
{\textit{Phys. Rev. A} \textbf{98}, 012324 (2018).}

\bibitem{Schuld2019_2}M. Schuld, V. Bergholm, C. Gogolin, J. Izaac and N. Killoran. 
{\textit{Phys. Rev. A} \textbf{99}, 032331 (2019).}

\end{thebibliography}


\clearpage
\renewcommand\thefigure{S\arabic{figure}}
\renewcommand\theequation{S\arabic{equation}}
\renewcommand\thetable{S\arabic{table}}

\setcounter{figure}{0}
\setcounter{equation}{0}
\setcounter{table}{0}

\renewcommand\thefigure{S\arabic{figure}}
\renewcommand\theequation{S\arabic{equation}}
\renewcommand\thetable{S\arabic{table}}
\renewcommand\thesection{\arabic{section}}
\renewcommand\thesubsection{\thesection.\arabic{subsection}}

%

\begin{center}
{\noindent {\bf Supplementary Material for\\
``Realizing a quantum generative adversarial network using a programmable superconducting processor''}}
\end{center}

%

\section{Details about the quantum device}
Our experimental device is a superconducting circuit consisting of 20 transmon qubits interconnected by a central bus resonator with frequency fixed at around $\omega_{R}/2\pi\approx$ 5.51 GHz. 
Five qubits, denoted as Q$_j$ for $j=0$ to 4, are actively used in this work.
Each qubit has its own microwave control and flux bias lines for implementations of XY and Z rotations respectively.
Meanwhile, each qubit is dispersively coupled to its own readout resonator for qubit-state measurement, and all the qubits can be measured simultaneously using the frequency-domain multiplexing technique.
Details about the fabrication process, circuits layout, and quantum manipulation of the device can be found in Ref.~\cite{song2017, song2019}.

\begin{figure}[htbp]
	\centering
	\includegraphics[width=3.4in,clip=True]{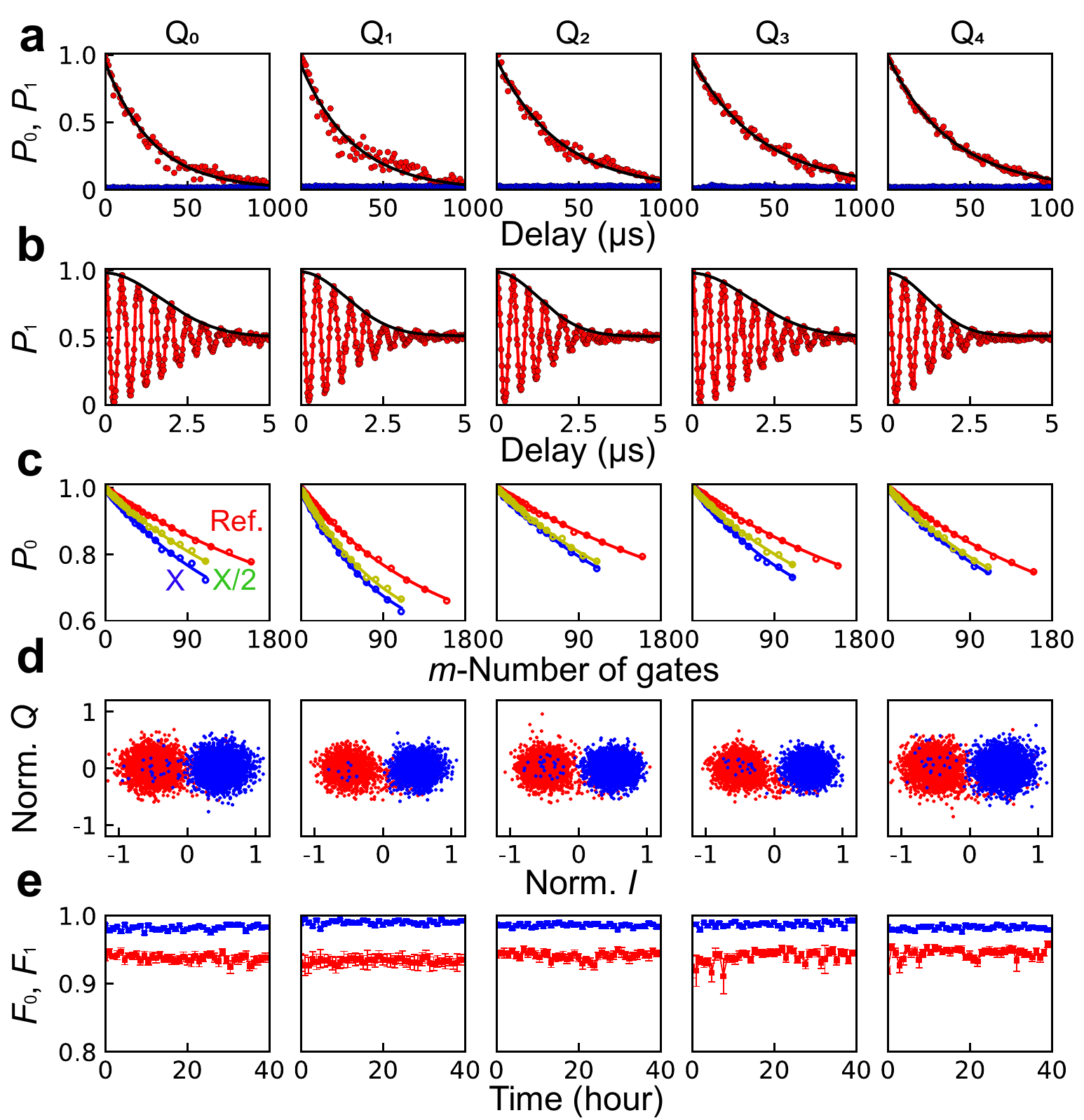}
	\caption{\textbf{Qubits characteristics.} \textbf{a,}~Qubit energy relaxation time measurement.
	The black solid lines are fits for $P_{1,j}$ of Q$_i$ according to the exponential decay equation, $P_{1,j}$ $\propto$ $exp(-t/T_{1,j})$. \textbf{b,}~Qubit Ramsey Gaussian dephasing time measurement.
	 The red dots are experimental data with its envelope of Ramsey fringe fitted by $P_{1,j}^{*}$ $\propto$ 0.5 + $exp[-t/2T_{1,j}-{(t/T_{2,j}^{*})}^2]$, shown by the black solid lines. \textbf{c,}~RB characterizing the fidelities of $X/2$ and $X$ gates. Plotted are the corrected $|0\rangle$-state probabilities of the five qubits averaged over $k$ = 30 sequences for the reference (red), the $X/2$ gates interleaved with the reference (green), and the $X$ gates interleaved with the reference (blue) as functions of the numbers of gates. Dots are experimental data and lines are fits by the exponential decay equation. \textbf{d,}~The measured normalized I-Q values in the I-Q plane, where blue (red) dots are obtained by repetitively preparing the qubit in $|0\rangle$ ($|1\rangle$) states and demodulating the corresponding readout signals. \textbf{e,}~The $|0\rangle$ and $|1\rangle$-state measurement fidelities of the five qubits with error bars, denoted respectively by $F_{0,j}$ (blue) and $F_{1,j}$(red) as a function of training time.
  \label{Exfig:qubitInfo}}
\end{figure}

Single-qubit operations and qubit-state measurements are performed at each qubit's idle frequency $\omega_{j,\textrm{idle}}$ as listed in Table~\ref{tab1}.
These frequencies are carefully arranged to minimize the crosstalk effects.
Other important parameters of the five qubit are also listed in Table~\ref{tab1}, with their corresponding benchmarking data shown in Fig.~\ref{Exfig:qubitInfo}. 
Both the $X$ and $X/2$ gates 
are designed following the DRAG theory~\cite{Motzoi2009} with a length of 30 ns and a full width half maximum of 15 ns.
Their fidelities are no less than 0.9976 characterized by randomized benchmarking (Fig.~\ref{Exfig:qubitInfo} (c)).
Repeated readout pulses, which are 1.5 $\mu s$ in length, are demodulated at room temperature, yielding the $I-Q$ points for each qubit on the complex plane forming two blobs to differentiate its states $|0\rangle$ and $|1\rangle$ (Fig.~\ref{Exfig:qubitInfo} (d)).
The probabilities of correctly reading out each qubit in $|0\rangle$ and $|1\rangle$ monitored over 40 hours are shown in Fig.~\ref{Exfig:qubitInfo} (e), with the corresponding mean values listed in Table~\ref{tab1}.

\begin{table}[!htbp]
	\renewcommand\arraystretch{1.5}
	\begin{tabular}{p{2.0cm}<{\centering} p{1.1cm}<{\centering}p{1.1cm}<{\centering}p{1.1cm}<{\centering}p{1.1cm}<{\centering}p{1.1cm}<{\centering}}
		\hline
		\hline
		Qubits  &  Q$_{0}$  &   Q$_{1}$  &   Q$_{2}$  &   Q$_{3}$  &   Q$_{4}$ \\
		\hline
		$\omega_{j,\textrm{idle}}/2\pi$ (GHz) & 5.220 & 5.055 & 5.105 & 5.165 & 5.000\\
		$T_{1,j}$ ($\mu s$) & 28.8 & 32.2 & 38.2 & 42.4 & 38.8 \\
		$T_{2,j}^{*}$ ($\mu s$) & 2.4 & 2.0 & 1.8 & 2.5 & 1.7  \\
		$X$/2 fidelity & 0.9992 & 0.9983 & 0.9990 & 0.9991 & 0.9992  \\
		$X$ fidelity & 0.9984 & 0.9976 & 0.9988 & 0.9979 & 0.9990  \\
		$F_{0,j}$ & 0.982 & 0.990 & 0.986 & 0.988 & 0.982  \\
		$F_{1,j}$ & 0.938 & 0.933 & 0.941 & 0.939 & 0.944  \\
		\hline
		\hline
	\end{tabular}
	\caption{\label{table1} \textbf{Qubits characteristics.} $\omega_{j,\textrm{idle}}$ is the idle frequency of Q$_j$ where single-qubit state preparation and measurement are implemented. Typical coherence parameters of the $j$-th qubit (Q$_j$), including energy relaxation time $T_{1,j}$ and Ramsey Gaussian dephasing time $T_{2,j}^{\ast}$ are measured at the idle frequency, respectively. Fidelities of the $X/2$ and $X$ gates on Q$_j$ are characterized by randomized
 benchmarking at its idle frequency $\omega_{j,\textrm{idle}}$. $F_{0,j}$ ($F_{1,j}$) is the typical $|0\rangle$ ($|1\rangle$)-state measurement fidelity for Q$_j$, which is then used to correct the measured qubit probabilities for eliminating the readout errors.}

\label{tab1}
\end{table}

\section{multi-qubit entangling gates}

The multi-qubit entangling gates that comprise \textbf{G} and \textbf{D} in our QGANs are generated by tuning all the involved qubits on-resonance at around 5.165 GHz, which is detuned from the resonator frequency by 345 MHz. Single-qubit phase gates, which are realized by amplitude-adjustable Z square pulses with a width of 20 ns, are added on each qubit before and after the interaction process to cancel out the dynamical phases accumulated during it. In our experiment, the two- and three- qubit entangling gates contain {Q$_{3}$, Q$_{4}$} and {Q$_{1}$, Q$_{2}$, Q$_{3}$} respectively. The characteristic interaction time $t_{gate}$ is fixed at $\pi/{4 |\lambda|}$, where $\lambda$ is negative and approximately equals to ${ \widetilde{g}}^{2}$/{$\Delta$}. $\widetilde{g}$ is defined as the average qubit-resonator coupling strength and $\Delta$ denotes the detuning between the interaction frequency and resonator frequency. $\lambda/2\pi$ are around -2.48 MHz and -2.27 MHz for the two- and three-qubit cases, respectively.

\begin{figure}[htbp]
	\centering
	\includegraphics[width=3.4in,clip=True]{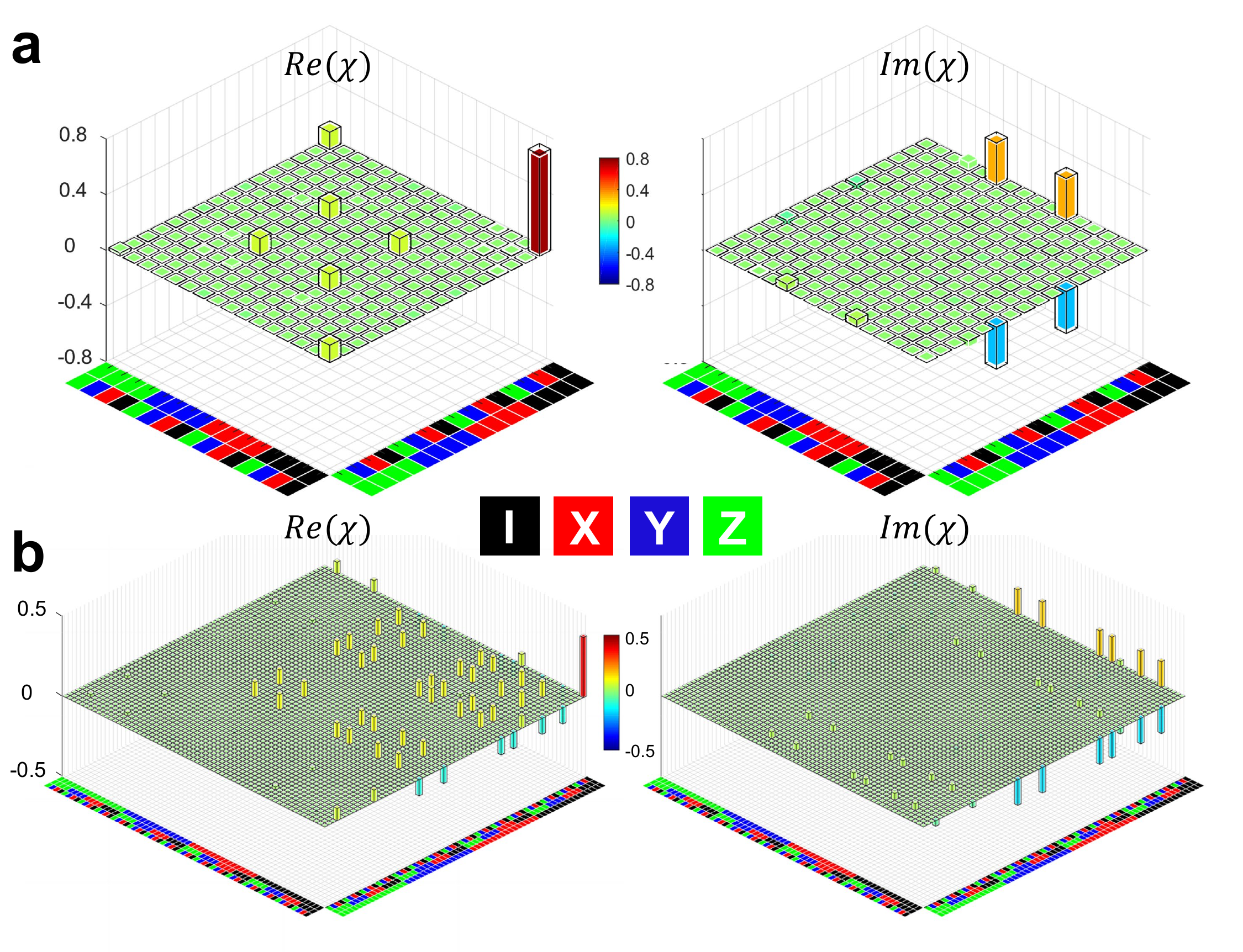}
	\caption{\textbf{Process matrices for entangling gates.} 
	The real and imaginary components of the experimental $\chi_{exp}$ (solid bars) and ideal $\chi_{ideal}$ (black frames) for the multi-qubit entangling gates are shown in the left and right panels.
	The fidelities are 0.9716$\pm$0.0110 and 0.9456$\pm$0.0154 for the two-qubit (\textbf{a}) and three-qubit (\textbf{b}) entangling gates, respectively.
  \label{Exfig:GHZchi}}
\end{figure}

To characterize the obtained gates, quantum process tomography (QPT) is performed by preparing a full set of $6^{n}$ input states ${\bigotimes_{j}^{n} \{I, \pm X/2, \pm Y/2,  X \}}$ (here $n$ denotes the number of qubits) and measuring the resulting output states with quantum state tomography (QST). We present the ideal and experimental process matrices $\chi_{\textrm{id}}$ and $\chi_{\textrm{exp}}$ for the two- and three-qubit entangling gates in Fig.~\ref{Exfig:GHZchi} (a) and (b), finding the gate fidelities tr($\chi_{\textrm{exp}}$$\chi_{\textrm{id}}$) of 0.9716$\pm$0.0110 and 0.9456$\pm$0.0154 respectively.

\begin{figure}[htbp]
	\begin{center}
		\includegraphics[width=3.4in,clip=True]{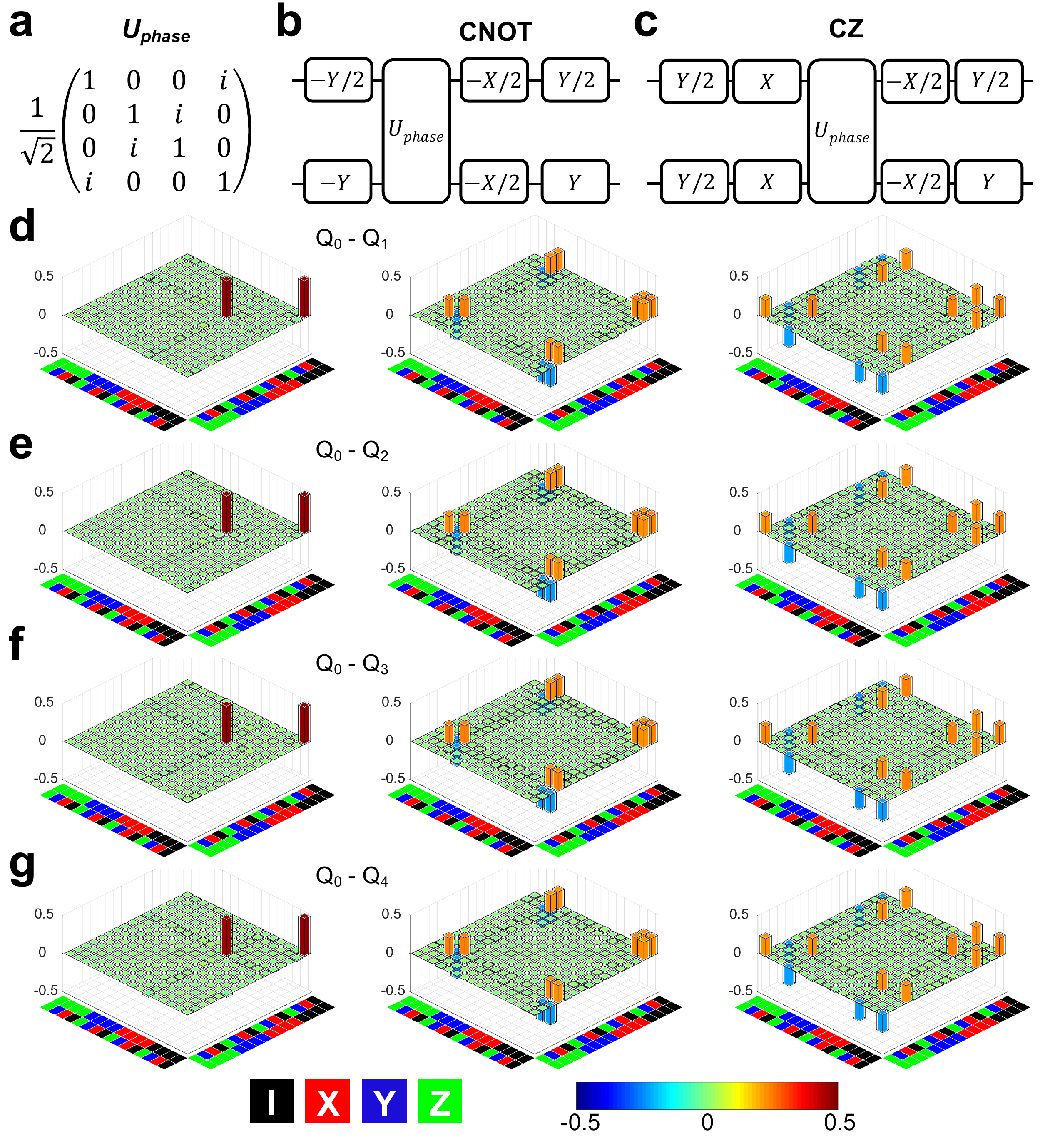}
		\caption{\textbf{Process matrices of two-qubit gates.}
		\textbf{a,}~The unitary matrix of the two-qubit  $U_{phase}$ gate. \textbf{b,}~The quantum circuit for constructing the CNOT gate, which consists of a $U_{phase}$ gate and six single-qubit gates. The first (above) qubit acts as the control qubit. \textbf{c,}~The quantum circuit for constructing the CZ gate, which consists of a $U_{phase}$ gate and eight single-qubit gates. The first qubit acts as the control qubit. \textbf{d,}~\textbf{e,}~\textbf{f,}~\textbf{\&} \textbf{g}~The real components of the process matrices $\chi_{exp}$ (solid bars) and ideal $\chi_{ideal}$ (black frames) of $U_{phase}$ (left), CNOT (middle) and CZ (right) gates obtained by QST for different pairs of qubits, i.e., Q$_{0}$-Q$_{1}$, Q$_{0}$-Q$_{2}$, Q$_{0}$-Q$_{3}$, and Q$_{0}$-Q$_{4}$. The first qubit in the pairs always acts as the control qubit. All process fidelities calculated by tr($\chi_{exp}$$\chi_{ideal}$) can be found in Table~\ref{tab2}. 
		}
		\label{Exfig:TwoQchi}
	\end{center}
\end{figure}

\section{Two-qubit controlled gates}

Two types of controlled gates, i.e., controlled-X (CNOT) and controlled-Z (CZ) gates, are frequently used to fulfill the quantum gradient subroutine in QGANs. Both gates are made from a two-qubit phase gate $U_{phase}$ whose unitary matrix is shown in Fig.~\ref{Exfig:TwoQchi} (a), combined with some single-qubit gates as shown in Fig.~\ref{Exfig:TwoQchi} (b) and (c). The experimental realization of the $U_{phase}$ gate can be found in Ref.~\cite{qguo2018}. For different pairs of qubits, the $U_{phase}$ gates have lengths of 159, 141, 170, and 162 ns respectively.
Ideal and experimental process matrices for all the related gates measured by QPT are shown in Fig.~\ref{Exfig:TwoQchi} (d) to (g), with the corresponding fidelities listed in Table~\ref{tab2}.

\begin{table*}[!htbp]
	\centering
	\renewcommand\arraystretch{1.5}
	\begin{tabular}{p{2.4cm}<{\centering}p{2.4cm}<{\centering}p{2.4cm}<{\centering}p{2.4cm}<{\centering}p{2.4cm}<{\centering}p{2.4cm}<{\centering}p{2.4cm}<{\centering}}
		\hline
		\hline
		Gate name & $\omega_{I}/2\pi$ (GHz) & $F_{U}$  & $F_{cnot}$  & $F_{cz}$ \\
		\hline
		Q$_{0}$-Q$_{1}$ & 5.050 & 0.9622$\pm$0.0059 & 0.9535$\pm$0.0062 & 0.9400$\pm$0.0061\\
		Q$_{0}$-Q$_{2}$ & 5.105 & 0.9760$\pm$0.0052 & 0.9446$\pm$0.0082 & 0.9499$\pm$0.0055\\
		Q$_{0}$-Q$_{3}$ & 5.165 & 0.9719$\pm$0.0102 & 0.9575$\pm$0.0049 & 0.9629$\pm$0.0048 \\
		Q$_{0}$-Q$_{4}$ & 5.000 & 0.9656$\pm$0.0065 & 0.9554$\pm$0.0058 & 0.9427$\pm$0.0099 \\
		\hline
		\hline
	\end{tabular}
	\caption{\label{table2} \textbf{Fidelities of two-qubit quantum logical gates.}  $\omega_{I}$ is the working point of the $U_{phase}$ gate. $F_{U}$, $F_{cnot}$, and $F_{cz}$ denote the process fidelities of $U_{phase}$, CNOT, and CZ gates obtained by QPT, respectively.}
\label{tab2}
\end{table*}

\section{The Gradient verification}

The gradient is a critical ingredient required in the optimization procedure of generative adversarial networks (GANs).
In this work, a quantum algorithm known as the Hadamard test is implemented experimentally for the first time to obtain the gradient.
Now we introduce the basic idea of this algorithm.

\begin{figure}[!htbp]
\begin{center}
		\includegraphics[width=3.4in,clip=True]{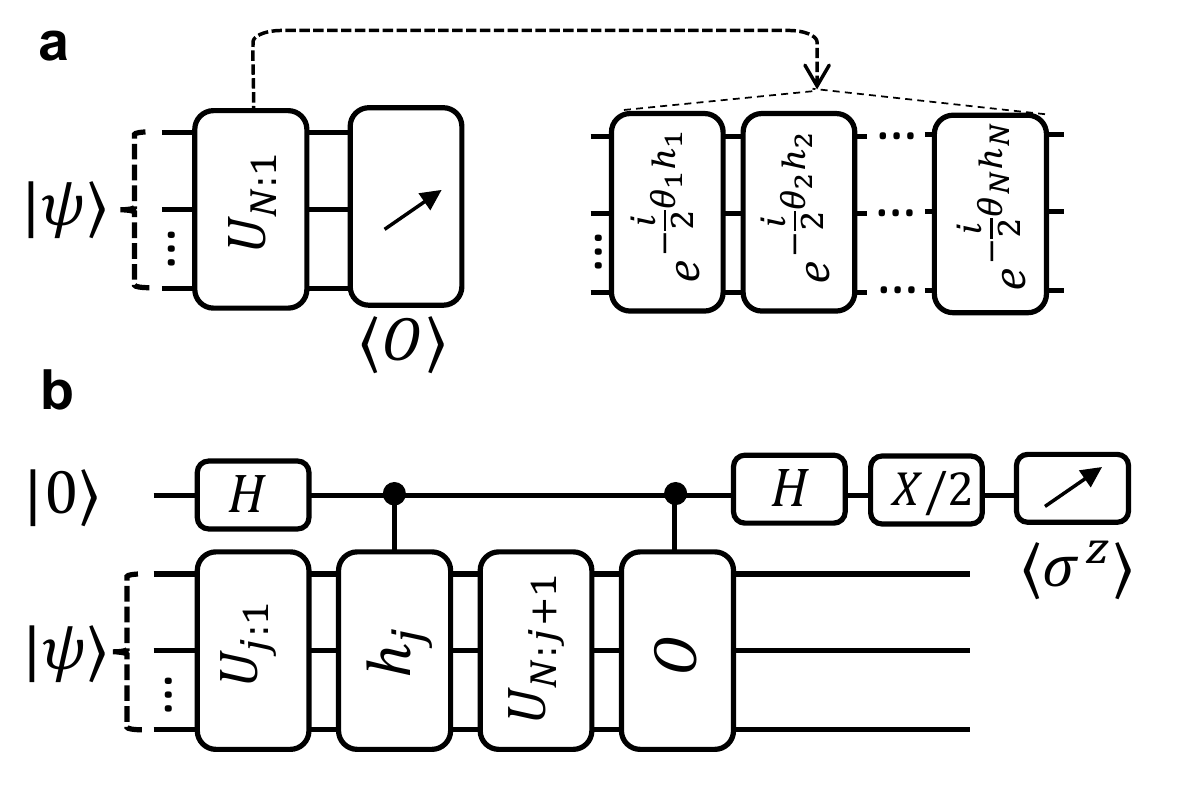}
		\caption{\label{fig:2}\textbf{Hadamard test.}
		\textbf{a,} Parameterized quantum circuit. The system consists of a bunch of qubits and is initialized in $|\psi\rangle$. $N$ unitaries, which are denoted as  $U_{N:1}=U_N U_{N-1} \cdots U_j \cdots U_2 U_1$, are applied to the system sequentially. Each unitary is parameterized by $U_j = e^{-\frac{i}{2}\theta_j h_j}$, where $h_j = h_j^\dagger$ is the Hamiltonian of the operation. The expectation value of an observable $O$ is measured at the end of the circuit.
		\textbf{b,} An instance circuit of Hadamard test, which is executed by introducing an ancillary qubit. To measure the gradient of $\langle O\rangle$ with respect to $\theta_j$, a controlled-$h_j$ gate is inserted after $U_j$, and a controlled-$O$ gate is appended at the end of the parameterized quantum circuit. The gradient is encoded in the expectation value $\langle\sigma^z\rangle$ of the ancillary qubit as explained in the text.}
	\label{Exfig:HadamardTest}
\end{center}
\end{figure}

A general parameterized quantum circuit for a multi-qubit system is depicted in Fig.~\ref{Exfig:HadamardTest} (a). The circuit consists of $N$ unitaries labeled as $U_j$ for $j=1$ to $N$ and outputs the expectation value of an observable $O$. $U_j$ can be written as $e^{-\frac{i}{2}\theta_j h_j}$ where $h_j = h_j^\dagger$ is the Hamiltonian of the system during the operation. 
Our goal is to measure $\nabla_{\vec{\theta}}\langle O\rangle$, i.e., the gradient of $\langle O\rangle$ with respect to $\vec{\theta}$, where $\vec{\theta} = [\theta_1, \theta_2, ..., \theta_j, ..., \theta_N]$.
According to the parameter shift rules, we can measure the analytic gradient with the `classical linear combinations of unitaries (CLCU)' method~\cite{Schuld2019_2} which requires the direct detection of $\langle O\rangle$.

Alternatively, we can add an ancillary qubit and measure the gradient indirectly by the Hadamard test algorithm.
An instance circuit of the Hadamard test to measure the partial derivative of $\langle O\rangle$ with respect to $\theta_j$ is shown in the circuit of Fig.~\ref{Exfig:HadamardTest} (b). For the input state $|0\rangle\otimes|\psi\rangle$, the corresponding output state of the circuit is
\begin{eqnarray}
&&\frac{1}{2}\left( |0\rangle\otimes (U_{N:1}+i O U_{N:j+1} h_j U_{j:1})|\psi\rangle \right. \nonumber \\
&&~~\left. +|1\rangle\otimes (U_{N:1}-i O U_{N:j+1} h_j U_{j:1})|\psi\rangle \right).
\end{eqnarray}
The expectation value of the ancillary qubit's $\sigma ^z$ is given by
\begin{eqnarray}
\langle \sigma ^z\rangle&=&\frac {i}{2}  \langle\psi|U_{1 ; j}^{\dagger}\left[U_{j+1: N}^{\dagger} O U_{N ; j+1}, h_{j}\right] U_{j: 1}|\psi\rangle \nonumber \\
&=&-\frac{\partial}{\partial \theta_{j}}\langle O\rangle
\label{Seq:gradient}
\end{eqnarray}
The proof of the last equation can be found in Ref.~\cite{NKpra2018}.
As such, the partial derivative of $\langle O\rangle$ can be measured as $-\langle \sigma ^z\rangle$ of the ancillary qubit.
In our experiment, since the parameterized unitaries are all single-qubit gates, only two-qubit controlled gates are required to obtain the quantum gradient (QG).
Nevertheless, the ability to realize two-qubit controlled gates between the ancillary qubit and all the other qubits in the system is still a nontrivial task and relies on the all-to-all connectivity of the quantum processor.

\begin{figure}[!htbp]
	\centering
	\includegraphics[width=3.4in,clip=True]{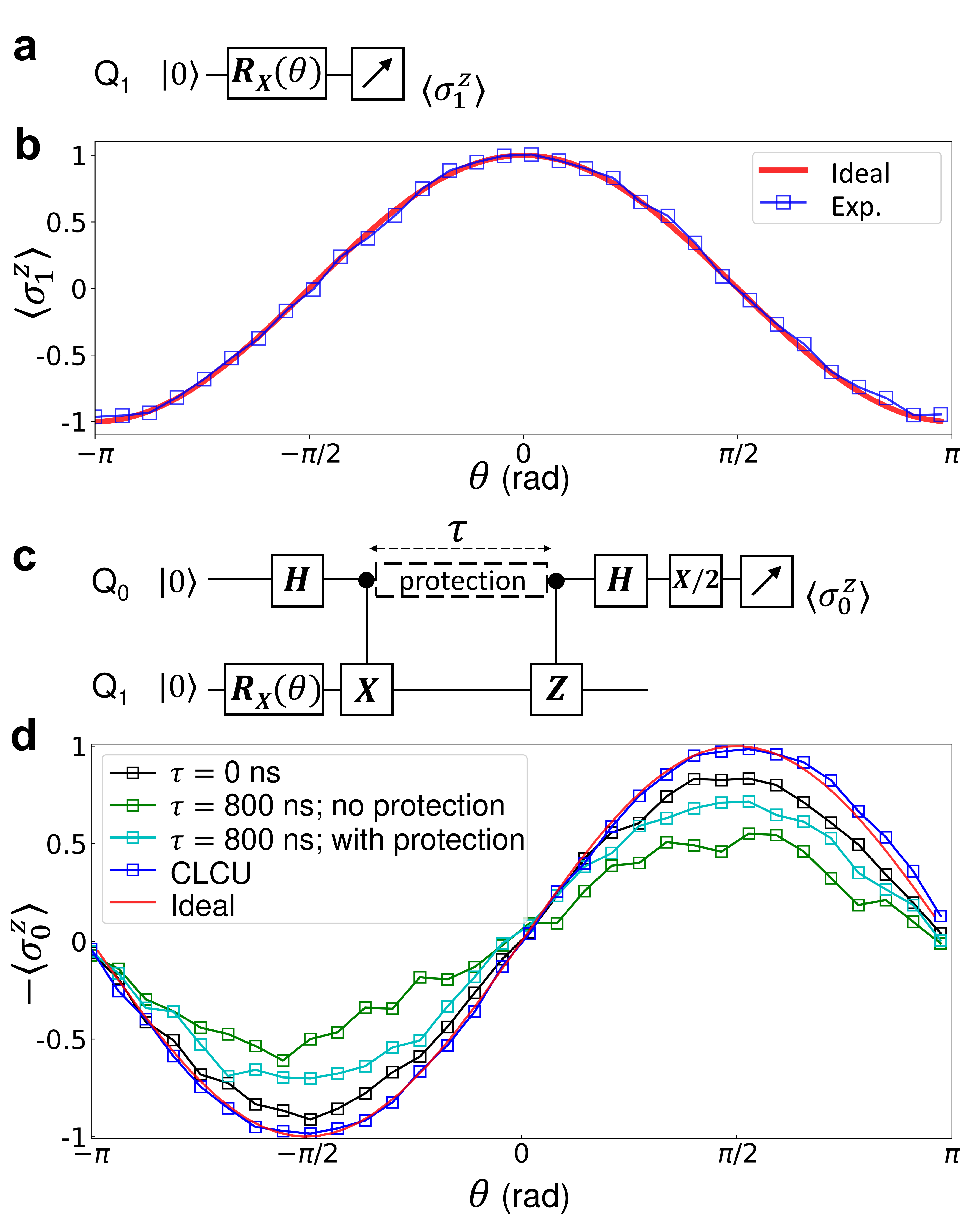}
	\caption{\textbf{Gradient verification.}
	\textbf{a,} A simple parameterized quantum circuit for testing QG.
	Our goal is to measure the gradient of $\langle\sigma_1^z\rangle$ with respect to $\theta$, i.e., $\partial\langle\sigma_1^z\rangle/\partial\theta$.
	\textbf{b,} Experimental (blue squares) and ideal (red line) results for running the parameterized quantum circuit with different $\theta$. The values of $\partial\langle\sigma_1^z\rangle/\partial\theta$ can be calculated by the classical linear combination of unitaries (CLCU)  or quantum gradient (QG) approach.
	\textbf{c,} The circuit for measuring gradient with QG approach.
	\textbf{d,}	Results for the gradients measured in different cases. Except for the ideal gradient (red line) and that calculated by the CLCU approach (blue squares), we also present the experimental results computed by the QG circuit for $\tau=$ 0 ns (black squares), $\tau=$ 800 ns without (green squares) and with (cyan squares) 1Q-DD protection respectively. 
}
	\label{Exfig:gradient}
	\centering
\end{figure}

We test the QG approach with a simple parameterized quantum circuit as shown in Fig.~\ref{Exfig:gradient} (a), where we first rotate Q$_1$ around the x-axis by an angle $\theta$ and then measure its $\langle\sigma_1^z\rangle$. Our goal is to measure the gradient of $\langle\sigma_1^z\rangle$ with respect to $\theta$, i.e., $\partial\langle\sigma_1^z\rangle/\partial\theta$. The variation of $\langle\sigma_1^z\rangle$ versus $\theta$ is directly measured, with the results shown in Fig.~\ref{Exfig:gradient} (b). Based on this measurement, we can calculate $\partial\langle\sigma_1^z\rangle/\partial\theta$ by the CLCU approach, and the result is shown in Fig.~\ref{Exfig:gradient} (d).
In comparison, we can also calculate the gradient by adopting the QG circuit as shown in Fig.~\ref{Exfig:gradient} (c) and measure $\langle\sigma_0^z\rangle$ of Q$_0$. The gradient can be obtained by $\partial\langle\sigma_1^z\rangle/\partial\theta = -\langle\sigma_0^z\rangle$ with the results shown in Fig.~\ref{Exfig:gradient} (d).
Due to the infidelities of our two-qubit controlled gates, the quantum gradient is slightly smaller than the ideal value.
In practice, there may be a delay between the two controlled gates which will introduce a dephasing error to Q$_0$ and affect the accuracy of QG.
Such an error can be reduced experimentally by the single-qubit dynamical decoupling (1Q-DD) technique~\cite{qguo2018}, which is realized by applying a continuous microwave drive resonantly on the qubit with a $\pi$-phase shift in the middle. 
To check that, we deliberately introduce a delay of $\tau=800$ ns in between the two controlled gates, and compare the results with and without the 1Q-DD protection. The Rabi frequency of the driving field in 1Q-DD is set to $\Omega/2\pi$ = 2 MHz in the experiment. An obvious enhancement is observed as expected.

\section{Pulse sequences for measuring the Quantum gradient}

\begin{figure}[!htbp]
	\centering
	\includegraphics[width=3.4in,clip=True]{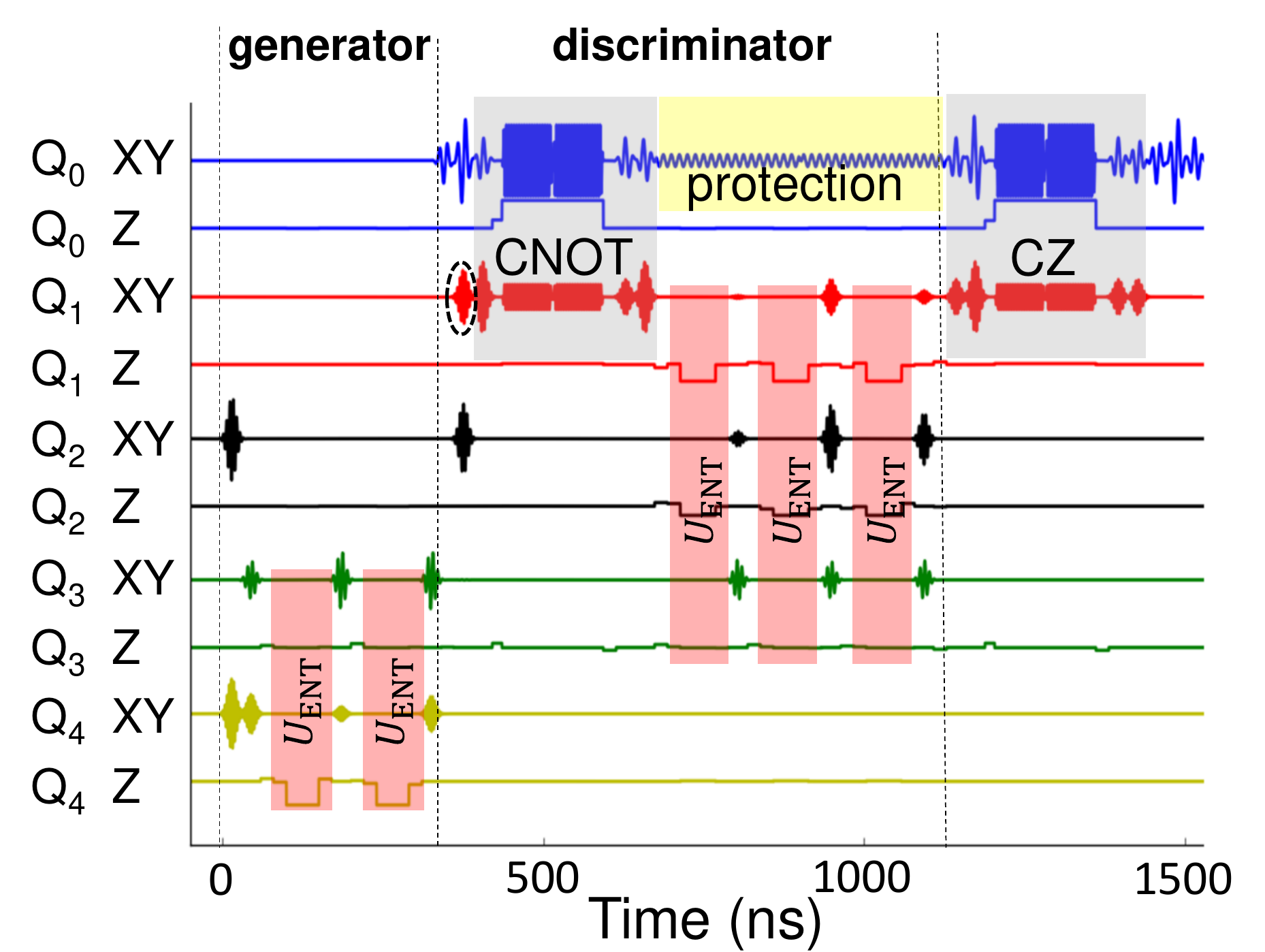}
	\caption{\textbf{Pulse sequence for measuring QG.} 
		Shown are the control signals of all five qubits for measuring the gradient of $\langle\sigma_1^z\rangle$ with respect to the angle of Q$_1$'s XY rotation, which corresponds to the pulse in the dashed circle.
		The two-qubit controlled gates, including CNOT and CZ, are located in the semitransparent blue blocks, while the two- and three-qubit entangling gates are covered by the semitransparent red shadows. In the yellow block, two weak microwave driving fields with opposite phases are applied successively to protect Q$_0$ from dephasing. The Rabi frequency of the driving field is set to $\Omega/2\pi$ = 2 MHz in the experiment. 
	}
	\label{Exfig:sequences}
	\centering
\end{figure}

The pulse sequence for measuring the gradient of $\langle\sigma_1^z\rangle$ to $\theta^x_{1,1,D}$ -- the first parameter of Q$_{1}$ in D -- during the training process of XOR gate is shown in Fig.~\ref{Exfig:sequences}. During the time between two controlled gates, we implement 1Q-DD technique on Q$_0$ to protect it from dephasing as introduced in the previous section. 

\section{Stability during training}

\begin{figure}[!h]
	\begin{flushleft}
	\includegraphics[width=3.4in,clip=True]{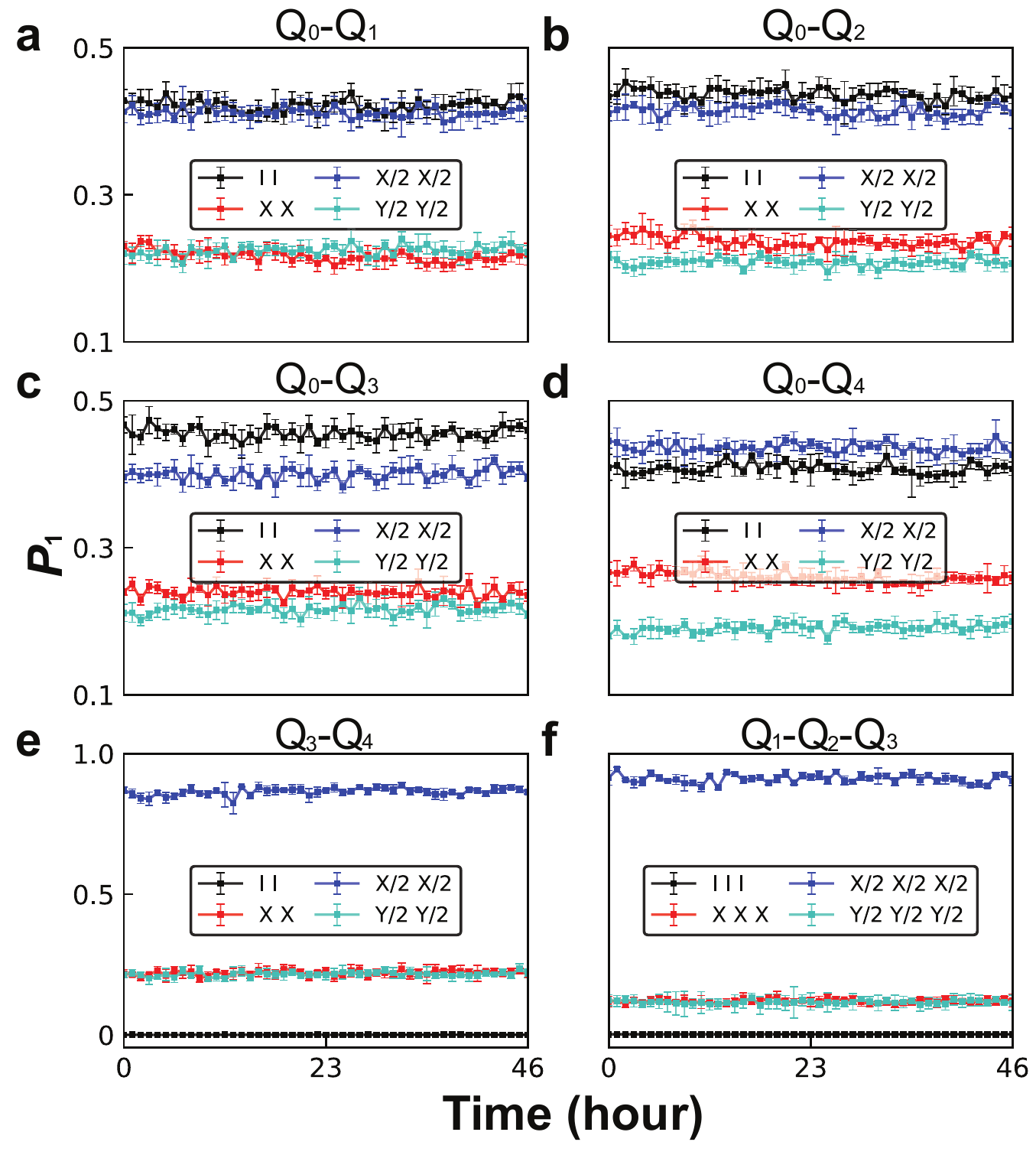}
	\caption{\textbf{Gates stabilities during the training period.}
		Shown are the experimental data for monitoring the stabilities of the two-qubit $U_{phase}$ gates (\textbf{a, b, c \& d}) and two entangling gates (\textbf{e \& f}) during a 40-hour training time. For each gate, we select four representative states as inputs and measure the probabilities of the output states with all qubits in $|1\rangle$. The input states are denoted by the operations that prepare them from the initial ground state.
			}
	\label{Exfig:gatefid}
	\end{flushleft}
\end{figure}

The stabilities of qubit operations are of great importance for the successful running of our QGANs. 
Except for the readout fidelities as shown in Fig.~\ref{Exfig:qubitInfo} (e), we also monitor the performances of multi-qubit gates during a 40-hour training time.
Instead of benchmarking the gates by QPT which is rather time-consuming, we select several representative states as inputs and monitor the corresponding output states by measuring the probabilities of all the qubits in $|1\rangle$.
The results for the $U_{\text{phase}}$ gates on four pairs of qubits,
i.e., Q$_{0}$-Q$_{1}$, Q$_{0}$-Q$_{2}$, Q$_{0}$-Q$_{3}$ and Q$_{0}$-Q$_{4}$, are presented in Fig.~\ref{Exfig:gatefid} (a)-(d). The variations of probabilities for all those gates are no more than 0.036. The results for two- and three-qubit entangling gates applied on Q$_{3}$-Q$_{4}$ and Q$_{1}$-Q$_{2}$-Q$_{3}$ are shown in Fig.~\ref{Exfig:gatefid} (e) and (f), with the variations of probabilities being 0.063 and 0.057 respectively.
All results show that the performance of our multi-qubit gates is sufficiently stable during the QGANs training.

\end{document}